\documentclass[aps,prb,twocolumn,amsmath,amssymb,showpacs,showkeys]{revtex4}
\usepackage{tabularx} 
\usepackage{graphicx} 
\usepackage{dcolumn} 
\usepackage{setspace}
\usepackage{bm}
\usepackage{wrapfig}                                                  
\usepackage{color}
\usepackage{fancyhdr} 
\usepackage{multirow}

\definecolor{Remarks}{rgb}{1,0.3,0.3}

\usepackage{floatflt}

\newcommand\COMMENTED[1] {}

\newcommand\GAUSSIAN[1][]{{\footnotesize{GAUSSIAN#1}}}

\newcommand\EWALD     {{\footnotesize{EWALD}}}
\newcommand\Ref[1]     {Ref.~\onlinecite{#1}}

\begin{document}

\title{
High sensitivity of $^{17}$O NMR to p-d hybridization in transition metal 
perovskites: 
first principles calculations of large anisotropic chemical shielding
}

\author{Daniel L. Pechkis, Eric J. Walter and Henry Krakauer} 
\affiliation{Department of Physics, College of William and Mary, Williamsburg, VA 23187-8795.}

\date{\today}

\begin{abstract}
A first principles embedded cluster approach is used to calculate
O chemical shielding tensors, $\hat{\sigma}$, 
in prototypical transition metal oxide ABO$_3$ perovskite crystals.
Our principal findings are 
1) a large anisotropy of $\hat{\sigma}$ between
deshielded $\sigma_{x}\simeq \sigma_{y}$ and shielded $\sigma_{z}$ components
($z$ along the Ti-O bond);
2) a nearly linear variation, across all the systems studied, of the isotropic 
$\sigma_\mathrm{iso}$ and uniaxial $\sigma_\mathrm{ax}$ components, as a function of the 
B-O-B bond asymmetry.
We show that the anisotropy and linear variation arise from large paramagnetic
contributions to $\sigma_{x}$ and $\sigma_{y}$ due to virtual transitions between
O(2p) and unoccupied B($n$d) states. 
The calculated 
isotropic $\delta_\mathrm{iso}$ and uniaxial $\delta_{\rm ax}$ chemical shifts
are in good
agreement with recent BaTiO$_3$ and SrTiO$_3$ single crystal $^{17}$O NMR measurements.
In PbTiO$_3$ and PbZrO$_3$, calculated 
$\delta_\mathrm{iso}$  
are also in good agreement with NMR 
powder spectrum measurements. 
In PbZrO$_3$, $\delta_\mathrm{iso}$ calculations of the five chemically distinct sites 
indicate a correction of the experimental assignments.
The strong dependence of $\hat \sigma$ on covalent O(2p)-B($n$d)
interactions seen in our calculations indicates that $^{17}$O NMR
spectroscopy, coupled with first principles calculations, can be 
an especially useful tool
to study the local structure in complex perovskite alloys.  

\end{abstract}

\pacs{71.15.-m,
        76.60.Cq 
        77.84.-s 
      31.15.Ar}

\keywords{Chemical shielding,
first principles,
NMR,
gaussian atomic basis,
GTO basis}

\maketitle

\section{Introduction} 

Due to their
reversible 
polarization 
and strong 
electro-mechanical coupling, 
ferroelectric perovskite 
oxides are key components in 
many electronic and mechanical devices such as sensors, actuators, and random 
access memory.~\cite{Ref:Scott_appl} 
The largest piezoelectric
response occurs when competing structural instabilities are present,
\cite{RabeBook} so that the local structure depends sensitively on strain, external fields,
and chemical composition, as 
for example, near the morphotropic phase boundary near $x=1/2$ in
Pb(Zr$_{1-x}$Ti$_x$)O$_3$ (PZT).
Nuclear magnetic resonance (NMR) can be an important probe of local structure,
and NMR has increasingly been used to study complex
ferroelectric alloys. 
\cite{ref:ZhouHoatsonVold-Pb,ref:ZhouHoatsonVold-Nb,ref:VijayakumarHoatsonVold-PMN} 
First-principles calculations of NMR properites are likely to play
an important role in helping to interpret complex measured spectra.

A material's characteristic NMR spectra is determined by the 
coupling of the nuclear magnetic dipole 
and electric quadrupole moments with the local
magnetic field and with the electric field gradients (EFG), respectively. \cite{CohenReif1957}
The chemical shielding tensor, $\hat{\sigma}$, determines the local magnetic
field at a nucleus 
%
\begin{equation}
\mathbf{B} = (1-\hat{\sigma})\mathbf{B}_{\rm ext},
\label{eq:B_ind}
\end{equation}
where the induced field $\mathbf{B}_{\rm ind}=-\hat{\sigma}\mathbf{B}_{\rm ext}$ 
arises from electronic screening 
currents. 
Theoretical determination of the chemical shielding tensor, $\hat{\sigma}$, is more
subtle than the EFGs, since the latter depend only on the ground state charge density.
Linear response  methods have long been used to calculate $\hat{\sigma}$
for isolated molecules and clusters. \cite{1999-Helgaker,H-termination,Si_Al-NMR-zeolite,N-NMR,Na-NMR,stueber:9236,EIM-Review} 
In extended systems complications arise due to the use of periodic boundary conditions (PBC),
\cite{ref:MPL,ref:GIPAW,2003-Profeta-SiO2}
and implementations have been limited to planewave-based methods.
Recently, an alternative to the linear response method has been proposed
based on calculations of the orbital magnetization.
\cite{thonhauser2005,niu2007,thonhauser2009,ceresoli2009}
If this approach proves successful, calculations of $\hat{\sigma}$ could be implemented
in other standard band structure codes, including all-electron methods. 
\cite{thonhauser2009}

Calculations of $\hat{\sigma}$ in crystals and extended systems have largely used
finite-size clusters, due to the lack of generally available
PBC computer codes with NMR functionality.
Accurate results can be obtained with this approach if the
screening currents are sufficiently localized near the target nucleus,
and if its local environment is adequately modeled by the 
cluster.  Embedded cluster techniques have been successfully used in
crystals and macromolecules to obtain nuclear quadrupole resonance
spectrum, ligand to metal charge transfer excitations, photoemission,
electric field gradients, and hyperfine coupling in high T$_c$
superconductors.  \cite{ref:martin_NQR,
  ref:martin_excitations,ref:Husser_hyperfine, ref:Renols_EFG_SC}

Such calculations have generally used standard all-electron quantum
chemistry methods, which are widely available in many quantum
chemistry computer programs,
such as {\GAUSSIAN}. \cite{Gaussian98,Gaussian03} These methods are very
mature and can calculate $\hat{\sigma}$ with a range of
approximations, using well tested Gaussian type orbital (GTO) basis
sets. \cite{EMSL_BasisSets2007} In increasing order of computational
cost, these methods range from mean-field type [density functional
theory (DFT), Hartree Fock (HF), and hybrid methods] to correlated
approximations [such as second-order Moller-Plesset (MP2) perturbation
and coupled cluster (CC)
methods]. \cite{1999-Helgaker,2004-bookkaupp,2007-Vaara} The present
cluster calculations exploit this flexibility to calculate chemical
shieldings, comparing DFT/GGA, restricted HF (RHF), and
hybrid-DFT/B3LYP calculations. Currently, only LDA and GGA DFT
approximations are available for NMR calculations in PBC methods.

For covalently bonded systems, cluster
calculations typically use hydrogen atoms to terminate dangling bonds
at the cluster surface.~\cite{H-termination} 
Instead, we use point
charge embedding \cite{ref:ewald} to model the long range Coulomb interactions in the
ABO$_3$ materials studied here. Additional techniques are also used to better handle the
polarizable character of the large O$^{2-}$ anion and also to control surface
depolarization effects. 

In this paper, we show that the embedded cluster approach can yield
accurate oxygen $\hat{\sigma}$ for transition metal perovskites.
The focus is on four prototypical materials: BaTiO$_3$,
SrTiO$_3$, PbTiO$_3$ and PbZrO$_3$.  Strong covalency between the O
and transition metal B atoms (and between the A and O atoms in Pb
based materials) delicately balance ionic electrostatic interactions
in these materials and related alloys, resulting in a wide variety of
interesting and technologically important properties.  The classic
ferroelectric material BaTiO$_3$ exhibits three reversible temperature
dependent ferroelectric phases.  SrTiO$_3$ is a key constituent in
superlattice structures 
with novel material properties.~\cite{RabeBook} PbTiO$_3$ is the
prototypical Pb based ferroelectric. The strong covalency of Pb on the
A-site is responsible for the large $c/a = 1.065$ ratio in PbTiO$_3$,
and it plays a similar role in large electromechanical response of
relaxor ferroelectrics such as PMN. \cite{ref:Pa97} PbZrO$_3$ adopts a
complicated non-polar antiferrodistortive structure with five
chemically inequivalent O sites, which our calculations reproduce.
PbZrO$_3$ is also of the end-point compounds, together with PbTiO$_3$,
of the PZT solid solution series.

The paper is organized as follows. The theoretical approach is
described in Sec.~\ref{sec:method}.
Calculated
results for ABO$_3$ systems are presented in Sec.~\ref{results}, and an
analysis of the results is presented in Sec.~\ref{discussion}. Finally, we 
summarize and conclude in Sec.~\ref{conclusions}.

\section{Theoretical Approach}
\label{sec:method} 

After briefly reviewing $\hat{\sigma}$ conventions, the construction
of embedded clusters for ionic materials is described.
Finally, we describe the quantum chemistry methods and GTO basis
sets used in the calculations.

\subsection{Chemical shielding tensor}
\label{sec:sigmatensor}

The chemical shielding tensor $\hat{\sigma}$ [Eq.~(\ref{eq:B_ind})] is
a mixed second derivative of the ground state energy with respect to
the applied magnetic field and nuclear magnetic moment.
\cite{1999-Helgaker} As such, it is an asymmetric second rank tensor
with nine independent components in general, although its symmetry can
be higher, depending on the site symmetry of the target
nucleus. \cite{sigma-symm-Buckingham} The anti-symmetric part of
$\hat{\sigma}$ contributes negligibly to the NMR resonance frequency
shift, since it enters only in second order,
\cite{Asymm-sigma1968,Asymm-sigma1991} although it can contribute to
relaxation. \cite{Asymm-sigma-relax} The symmetric part can always be
diagonalized, and the NMR frequency is determined by the following
combinations of its principal axis components \cite{baugher1969}
\begin{equation}
\begin{array}{c}
 \sigma _{{\rm{iso}}}  = {1 \over 3} \left( {\sigma _x  + \sigma _y  + \sigma _z } \right)  
                       = {1 \over 3} \rm{Tr}\,\hat{\sigma}\\ 
 \sigma _{{\rm{ax}}}  =  {1 \over 6} \left( {2\sigma _z  - \sigma _x  - \sigma _y } \right)
                       = {1 \over 2} \left( {\sigma _z - \sigma _{{\rm{iso}}} } \right)   \\ 
 \sigma _{{\rm{aniso}}}= {1 \over 2}\left( {\sigma _y  - \sigma_x        } \right), \\ 
 \end{array}
\label{eq:sigmaiso}
\end{equation}
where $\sigma _{{\rm{iso}}}$, $\sigma _{{\rm{ax}}}$, and $\sigma _{{\rm{aniso}}}$ are the 
isotropic, uniaxial, and anisotropic components, respectively.
In this paper, $\sigma _z$ is chosen to correspond
to the principal axis that is  mostly nearly along the \mbox{B-O-B} bond direction.

Positive values of $\sigma$ are conventionally denoted as shielding 
the external field, while negative elements are referred to as deshielding [see Eq.~(\ref{eq:B_ind})].
Measurements of $\hat{\sigma}$ are usually reported with
with respect to a
reference compound, where the chemical shift tensor is defined as~\cite{2004-bookkaupp}
\begin{equation}
\hat{\delta} =  -(\hat{\sigma}  - \sigma_\mathrm{iso}^\mathrm{ref}).
\label{eq:delta}
\end{equation}

\subsection{Embedded clusters}
\label{sec:cluster}

\begin{figure*}[p]
\begin{center}
\includegraphics[scale=0.85,clip=no]{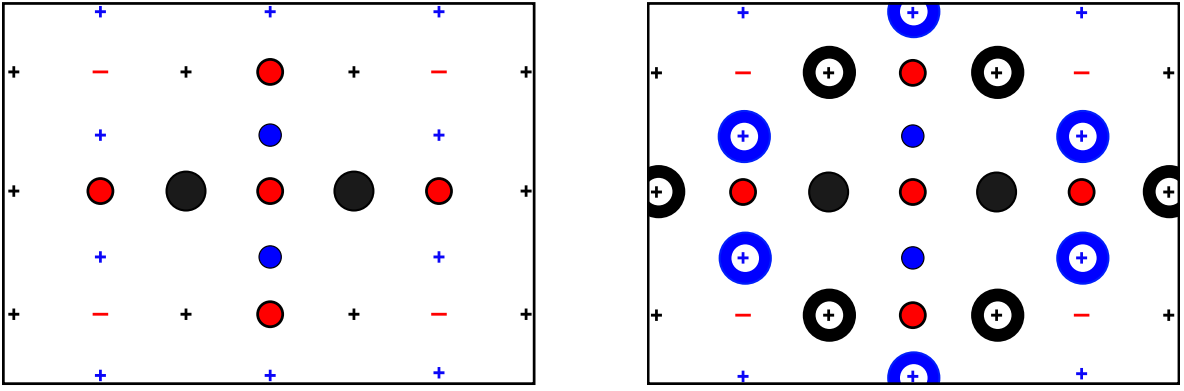}
\caption{Illustration of cluster embedding for the O-centered
A$_4$B$_2$O$_{15}$ 
quantum atom (QM) cluster (Sec.~\ref{QM_cluster}),
shown for a [110] plane of the ideal perovskite structure. 
The QM A, B, and O atoms are depicted by filled 
black, blue and red circles, respectively.
In the left panel, the QM atoms are embedded in point charges, ``$+$'' or ``$-$'' signs
colored coded according to the QM atom they replace in the crystalline lattice.
In the right panel, the boundary O QM atoms have had their nearest and next-nearest neighbor
cation point charges replaced by ``empty'' pseudopotentials (Sec.~\ref{ePSP}), indicated by
thick large circles, with corresponding color coding.}
\label{fig:embedded} 
\end{center}
\end{figure*}

An embedded cluster model is depicted in Fig.~\ref{fig:embedded}.
Its core consists of real, or ``quantum'' (QM) atoms, with the target
O atom at its center.
The total number of electrons is determined by
satisfying the nominal formal valence of all the QM atoms.  
The A$_4$B$_2$O$_{15}$ QM cluster in the figure thus has a net charge 
of -14.
The QM cluster is embedded in the classical potential
due to point charges, pseudopotentials, and, in some cases, an external electric field to cancel
surface depolarization effects, as described in 
following subsections.
The atomic site designations nn and nnn, used below, denote
nearest-neighbor and next-nearest-neighbor sites, respectively, 
and are
based on the ideal structures. Thus, for example, the Ti atom in tetragonal P4mm
PbTiO$_3$ would be regarded as having six nn O atoms, despite the distortion
of the TiO$_6$ octahedron. 
In the actual calculations, the tetragonal distortion results in 
two chemically inequivalent O$_{\rm eq}$ 
and O$_{\rm ax}$ atoms (O$_{\rm ax}$ lying on the polar axis),
which require two separate embedded cluster calculations.

\subsubsection{Quantum cluster}
\label{QM_cluster}

In all the QM clusters considered in this paper, the target oxygen atom is fully
coordinated with QM atoms located at its nn and nnn sites. Secondly,
the target atom's nn QM atoms are themselves fully coordinated with
nn QM atoms. Finally additional QM atoms are added, as required by ideal perovskite symmetry. 
This procedure results in a 21 QM-atom cluster:
(A$_4$B$_2$O$_{15}$)$^{14-}$, depicted in  Fig.~\ref{fig:embedded}, where A = Sr, Ba, or Pb; B = Ti or Zr. 
This includes two corner-shared BO$_6$ octahedra, centered on the targeted O atom
(11 O and 2 B atoms), the target atom's 4 nnn A cations, 
and 4 additional O atoms, which are also one lattice constant away from the center atom.
Tests with larger QM clusters were used to estimate convergence of the
chemical shielding tensor with respect to cluster size, as discussed in Sec.~\ref{sec:QMchemistry}.

\subsubsection{Madelung potential: point charges}
\label{Madelung}

Next, the QM cluster is embedded in the crystal environment by
surrounding it with a large array of point charges.  The purpose
of the point charges is to better simulate the crystal environment by
generating the correct crystalline electrostatic Madelung
potential in the QM region.
The correct Madelung potential also plays a key role in stabilizing
the O$^{2-}$ ion, as has been shown in Gordon-Kim \cite{gordonkim} type models, \cite{PIB1986,PIB1987}
where
the internal energy of ionic systems is determined from the
energy of overlapping ionic charge densities. 
The finite point charge 
distribution is determined using 
the {\EWALD} program,  \cite{ref:ewald} 
as follows.
In a first step, {\EWALD} calculates the Madelung potential with the Ewald method for PBC,
using nominal ionic values  
({\em e.g.}, $\tilde{Q_i}=-2$ and $\tilde{Q_i}=+2$ for O$^{2-}$ and
Pb$^{2+}$, respectively) for the atoms placed at crystallographic positions of 
the targeted system.
In a second step, {\EWALD} retains the nearest ${\mathcal O}(10^4)$ $\tilde{Q_i}$ centered
on the target atom, adjusting the values of the outermost $\tilde{Q_i}$
to reproduce the Madelung electrostatic potential on and in the vicinity of the 
QM atoms.  
In this second step,
the nearest $\simeq 500-750$ $\tilde{Q_i}$ are fixed at their nominal values, and, in addition, the
net monopole and dipole moments of the point charge distribution are constrained to vanish.

\subsubsection{Boundary compensation via empty pseudopotentials }
\label{ePSP}
To accelerate the convergence of $\hat{\sigma}$ with respect to the size
of the QM cluster, it is advantageous to control 
the artificial polarization of boundary O(2p) states.
This arises  from the strongly attractive 
electrostatic potential of neighboring cation point charges for QM oxygen 
atoms on the periphery
of the cluster.
To alleviate this, the nn and nnn cation point charges of boundary O atoms are replaced
by ``empty'' pseudopotentials (ePSP), \cite{ref:TIPs} 
as illustrated in Fig.~\ref{fig:embedded}.
The ePSP is defined as follows: 
i) it is a large core pseudopotential for the most loosely bound valence electrons
of the corresponding cation ({\em e.g.}, a Ti$^{4+}$ PSP); ii) there are no
GTO basis functions associated with this site. 
The resulting modified cation classical potential simulates the
Pauli cation core repulsion, thereby
reducing the artificial polarization of boundary O(2p) states. \cite{ref:TIPs}
In the embedded 21 atom QM cluster discussed above, for example,
26 boundary point charges are replaced by ePSPs (labeled as A$^*$ or B$^*$),
yielding a (A$_4$B$_2$O$_{15}$)$^{14-}$ - A$^*_{16}$B$^*_{10}$ cluster, which is surrounded by the
remaining $\simeq {\mathcal O}(10^4)$ point charges.

\subsubsection{Cancellation of electric depolarization fields}
\label{polarcluster}
 
Due to the spontaneous electric polarization found in some materials, such as
tetragonal PbTiO$_3$, a macroscopic depolarizing electric field is
present in finite samples, in the absence of surface compensating charges.
Calculations using PBC are implicitly done in zero total macroscopic electric field, 
which automatically excludes surface depolarization effects.
In the present finite size clusters, a depolarizing electric field can arise from
a possible net dipole moment, due to polarization of the quantum mechanical charge density, 
{\em i.e.}, from the wave functions of the QM cluster.
The net dipole moment due to the point charges and ePSPs is zero by construction,
as discussed above.
The resulting depolarizing electric field
is removed in the calculation,
by applying an external electric field in the opposite
direction. The magnitude of the external field is chosen so that the force on the
central target atom matches that of
an all-electron PBC linearized
augmented planewave (LAPW) \cite{ref:LAPW} calculation. In normal equilibrium conditions,
using experimentally determined structures, 
the LAPW forces are usually small, since the theoretical structure is usually close to that of experiment.


\subsection{Calculational details}
\label{sec:QMchemistry}

The embedded cluster calculations were performed with the {\GAUSSIAN} computational 
package. \cite{Gaussian98,Gaussian03}
The chemical shielding tensor was determined using the continuous 
set of gauge transformations (CSGT) method. \cite{ref:CSGT,cheeseman} 
The gauge-independent atomic orbital (GIAO) 
method~\cite{Ditchfield1974} was also used for comparison in some cases. 
The gauge origin in GIAO is at the target atom, yielding
a useful decomposition into diamagnetic and paramagnetic components,
as discussed further below and in the next Section.
We did not include a surface dependent magnetic depolarization contribution
to $\delta_\mathrm{iso}$, which
depends on the magnetic susceptibility.
\cite{NMRconventions_2008,magsuscep_foot}

Calculations were done using the DFT hybrid B3LYP, \cite{B3LYP} as
well as the generalized gradient approximation (GGA), using the PW91
form. \cite{ref:PW91} The B3LYP calculations were found to yield
generally better agreement with experiment (Sec.~\ref{results}) than
GGA.  Site-centered GTO basis functions were associated with all the
QM atoms. All-electron treatments were used for the O and Ti atoms,
while the other QM atoms were represented using scalar-relativistic
small core (RSC) pseudopotentials [also called effective core
potentials (ECPs)]. All basis sets and ECPs listed below were taken
from the EMSL website. \cite{EMSL_BasisSets2007} O-centered
(A$_4$B$_2$O$_{15}$)$^{14-}$ QM clusters were used in the
calculations, suitably embedded with point charges and ePSPs, as
discussed in Sec.~\ref{sec:cluster}).  Test calculations for larger
clusters were used to check size convergence, as discussed at the end of this
section.

The RSC ECPs for the QM atoms were
Sr(28), Zr(28), Ba(46), Pb (60), where the number of core electrons is
shown in parenthesis.
These pseudopotentials are generally specified by the same label as their associated basis
sets listed below, except where otherwise indicated.
For ePSPs (Sec.~\ref{ePSP}), which have no associated GTOs,
scalar-relativistic large core (RLC) ECPs were used as follows:
CRENBS RLC for Ti(18) and Zr(36); Stuttgart RLC for Sr(36) and Ba(54).
[The Ba ePSP was also used for the Pb ePSP atoms, because we
could not find a large core Pb$^{2+}$ quantum chemistry type pseudopotential
in the standard databases.]  The following
GTO basis sets were found to give well-converged results: 
O(IGLO-III), Ti(cc-pwCVTZ-NR), Zr(cc-pwCVTZ-PP), and Pb(cc-pVTZ-PP).
For the Sr and Ba atoms, the associated Stuttgart RSC 1997 basis sets were used.
The correlation consistent basis sets were employed
in this study, because this basis function series facilitates
a systematic study of chemical shift convergence with respect to basis set size. 
(We could not find correlation consistent basis sets for Sr and Ba.)
The O(IGLO-III) basis set was used, because it was specifically designed for magnetic 
property calculations. In convergence tests, we found that it had quadruple-$\zeta$
(QZ) accuracy even though it has fewer basis functions. 

In SrTiO$_3$, for example, B3LYP CSGT  principle
value components
calculated with the aforementioned basis sets
differed by no more than 10 ppm from results calculated with 
the O(cc-pwCVQZ), Ti (cc-pwCVQZ-NR),
and Sr (def2-QZVP) basis sets,
while $\delta_{\mathrm{iso}}$ differed by only $\simeq 8$~ppm. 
In PbTiO$_3$, 
changing the Pb basis set from DZ to TZ changed 
$\delta_{\mathrm{iso}}$ by less than $10$~ppm, while
the difference between inequivalent O sites changed by less than 4
ppm. These results are consistent with calculations for SrTiO$_3$, 
using two different gauge methods, shown in Table~\ref{tab:O-tensors},
where, in the infinite basis set limit, the CSGT and GIAO values should agree.
The individual B3LYP components, $\sigma_x = \sigma_y$ and $\sigma_z$, are seen to
differ by 3 and 26 ppm, respectively, while $\sigma_{\mathrm{iso}}$ differs
by only 6 ppm.

Size convergence errors of about 30 ppm in the absolute value of the chemical shieldings
were estimated from calculations on larger clusters, using the above basis sets.
In SrTiO$_3$, for example, increasing the O-centered  cluster size from the
Sr$_4$Ti$_2$O$_{15}$ (21 atoms) to Sr$_{4}$Ti$_{10}$O$_{47}$ (61 atoms)
changed the RHF, B3LYP, and PW91 $\sigma_\mathrm{iso}$ by -13, -27, and -25 ppm, 
respectively, while individual principal values changed by less than 16, 30, and 37 ppm. 
These test calculations used large core Ti pseudopotentials for the additional Ti atoms.

\section{Results}
\label{results}

Calculated  chemical shielding results 
for the prototypical perovskites SrTiO$_3$, BaTiO$_3$, PbTiO$_3$ and PbZrO$_3$
are presented and compared to $^{17}$O NMR single crystal and powder
 spectra chemical shift measurements in this section.
Calculations were performed for embedded clusters corresponding to the
following structural parameters.  For cubic SrTiO$_3$, the lattice
parameter of \Ref{ST_a} was used.  BaTiO$_3$ cubic and tetragonal P4mm
structures were taken from \Ref{ref:BT_phasestruct}; rhombohedral R3m
from \Ref{ref:BT_r3m}. The PbTiO$_3$ tetragonal P4mm structure was
taken from \Ref{ref:NelmsKuhs}. For PbZrO$_3$,
which has a complicated Pbam unit cell containing eight formula units,
experimental lattice
parameters from neutron scattering measurements
were used together with internal coordinates determined from first
principles calculations. ~\cite{ref:Johannes}  All calculations were carried out using the
embedded cluster approach described in Sec.~\ref{sec:method}, using
the hybrid B3LYP exchange-correlation functional. In some cases RHF
and GGA/PW91 results are reported for comparison.

\label{sec:results-O} 

Table~\ref{tab:O-tensors} presents our calculated results for the principal values
of the symmetrized chemical shielding tensor $\hat{\sigma}$. (The asymmetry in the 
chemical shielding tensors was less than 0.5 ppm; see Sec.~\ref{sec:sigmatensor}.)
The principal axis frame of the target O atom is indicated by $x$, $y$, $z$, where
the $z$-axis is always identified with the local \mbox{B-O-B} direction. This is 
exact in the cubic and tetragonal crystals. For 
lower O-site symmetry, the $z$-axis well approximates the quasilinear 
\mbox{B-O-B} bond in all the structures considered here.

A striking feature in Table~\ref{tab:O-tensors} is
the large $\sigma_{x}\simeq \sigma_{y}$ vs. $\sigma_z$ anisotropy of the principal values for each case,
with 
two large negative (deshielded) principal values and one considerably smaller positive 
(shielded) $\sigma_z$ principal value.
The O atoms have their highest site-symmetry in the cubic crystals, 
resulting in an exact two-fold $\sigma_x=\sigma_y$ degeneracy.
For lower symmetry cases,
the degeneracy is lifted, but the splitting remains small in most cases; the 
largest splitting is 65 ppm for the O1-4g site in PbZrO$_3$.
As shown by the GIAO calculations for SrTiO$_3$, the anisotropy arises from the paramagnetic
components. 
We note that the GIAO diamagnetic components in SrTiO$_3$ differ by only $\simeq 40 - 50$~ppm from that of the
the isolated, purely diamagnetic, O$^{2-}$ atom. 

In the fake SrTiO$_3$ calculation, where
d-like ($\ell=2$) Ti-centered GTO basis functions were deliberately excluded from the calculation,
the large $x,y$ vs. $z$ anisotropy is absent. The fake diamagnetic values are closer to the atomic O$^{2-}$ 
shielding, and the paramagnetic contributions are close to isotropic.
The $x,y$ vs. $z$ anisotropy of the calculated principal values 
and its relation to O(2p) hybridization with the B-atom d-states
is analyzed in Sec.~\ref{sec:p-d}.

Table~\ref{tab:ST-BTshift} compares the results for SrTiO$_3$ and BaTiO$_3$ in 
Table~\ref{tab:O-tensors} with recent
single crystal measurements \cite{ref:BT-ST-O_Blinc} and with earlier powder spectrum measurements
of the isotropic shifts.
Isotropic and uniaxial components 
of the chemical shift tensors are given, where 
$\delta_{\mathrm{ax}} = (\delta_{z} - \delta_{\mathrm{iso}})/2 $, using Eq.~(\ref{eq:sigmaiso}). 
In SrTiO$_3$, the calculated $\delta_{\mathrm{iso}}$ and
$\delta_{\mathrm{ax}}$ differ from experiment by $\simeq$ 20 ppm and
$\simeq$ 9 ppm respectively.  In BaTiO$_3$, $\delta_{\mathrm{iso}}$
calculated and experiment values differ by at most 16 ppm.

Table~\ref{tab:PT-PZshift} compares the results for PbTiO$_3$ and PbZrO$_3$ 
isotropic shifts $\delta_{\mathrm{iso}}$ (derived from Table~\ref{tab:O-tensors}) 
with well resolved measured powder spectra.~\cite{ref:Baldwin05}
For both systems, the calculated results differ from experiment by no
more than 21 ppm, with an average discrepancy of approximately 10 ppm.
Differences of relative shifts between chemical sites are smaller in most cases.
For example, in tetragonal PbTiO$_3$ the calculated B3LYP splitting
between the two inequivalent O sites is $\simeq 200$ ppm, in good
agreement with measured powder spectra.~\cite{ref:Baldwin05}

\section{Discussion}
\label{discussion}

A key feature in the electronic structure of transition metal perovskites
are the covalent interactions
between the O(2p) and the (formally) unoccupied transition metal-d
states. Indeed, the delicate balance between covalent and
electrostatic ionic interactions is responsible for the wide variety
of interesting properties exhibited by these materials and related
alloys.  In this section, we analyze the calculated chemical
shieldings in Sec.~\ref{results} 
in relation to p-d hybridization. 


The measured NMR $^{17}$O spectra show narrow well separated peaks in these materials, 
indicating that second-order quadrupolar broadening and, thus, the electric field gradients (EFGs)
are small, \cite{ref:Baldwin05} consistent with first principles calculations of O EFGs. 
\cite{ref:Johannes,ref:mao014105}
For the systems studied here, all the O sites are clearly resolved in the measurements,
and the chemical shielding tensor largely determines the second order 
quadrupolar peak positions. \cite{ref:Baldwin05}

\subsection{p-d hybridization and anisotropy of oxygen chemical shielding}
\label{sec:p-d}

As mentioned, a striking feature in Table~\ref{tab:O-tensors} is the 
large anisotropy of the chemical shielding principal values.
This is due to hybridization between the O(2p) and virtual B-site d-states.
The qualitative features can be understood from a simplified picture, 
which focuses on the \mbox{B-O-B} quasilinear structural unit (Fig.~\ref{fig:embedded}).
In a linear molecule, the dependence 
of $\hat{\sigma}$ on the direction of the applied field $\mathbf{B_\mathrm{ext}}$
is particularly simple, if we choose to locate the
vector potential gauge origin at the
target nucleus. \cite{Ramsey1950,Ramsey1951} 
In this case, when $\mathbf{B_\mathrm{ext}}$
is parallel to the molecular axis, only the diamagnetic (shielding) component
contributes to $\hat{\sigma}$; when $\mathbf{B_\mathrm{ext}}$ is perpendicular to the
axis, both diamagnetic and paramagnetic (deshielding) components contribute.
(Note that the calculated $\hat{\sigma}$ are invariant under a change of
the gauge origin. Consequently, 
we are free to interpret the $\hat{\sigma}$ values as
if they had been calculated with any particular choice of origin.)
The diamagnetic component depends only on ground state wave functions, while
the paramagnetic component depends on virtual transitions to unoccupied states, {\em i.e.,}
occupation of virtual states in the first order perturbed wave functions. 
In a DFT or HF calculation, this implies that large paramagnetic contributions 
could be expected,
if there are low lying unoccupied one particle eigenstates that are strongly coupled to 
occupied states. This is the case in these materials, where there is
strong coupling between O(2p) and virtual B-site d-states.

In cubic SrTiO$_3$, for example, when $\mathbf{B_\mathrm{ext}}$ is
applied along the Ti-O bond direction ($z$-direction),
the O nucleus is shielded by the applied field ({\em i.e.,} the principal value $\sigma_{z}=90$ is positive 
[Eq.~(\ref{eq:B_ind})]).  By contrast, when $\mathbf{B_\mathrm{ext}}$ is perpendicular to the Ti-O bond, 
the O nucleus is strongly deshielded ($\sigma_{x}=\sigma_{y}=-343$) as shown in Table~\ref{tab:O-tensors}.
According to this \mbox{B-O-B} picture, we would
infer that only diamagnetic contributions should contribute to $\sigma_{z}=90$,
and the positive $\sigma_{z}=90$ value is consistent with this. Similarly,
we would attribute the deshielded $\sigma_{x}=\sigma_{y}=-343$ value to
paramagnetic contributions arising from the O(2p)-Ti(3d) hybridization mechanism.
This interpretation is supported by the 
fake SrTiO$_3$ calculations, where removal of Ti d basis functions quenches
O(2p)-Ti(3d) hybridization and results in nearly isotropic principal value
components, $\sigma_x$ = $\sigma_y$ $\simeq$ $\sigma_z$.

While qualitatively correct, the simple \mbox{B-O-B} model leaves out important crystalline effects.
Thus the SrTiO$_3$ GIAO total shielding value $\sigma_{z}=116$ arises from cancellation
of large diamagnetic $\sigma_{\mathrm{d},z}=371$ and smaller paramagnetic 
$\sigma_{\mathrm{p},z}=-255$ components. 
Note that the fake $\sigma_{\mathrm{p},z}=-262$ value is similar to the normal
$\sigma_{\mathrm{p},z}=-255$ value; in addition 
$\sigma_{\mathrm{p},xy}\simeq\sigma_{\mathrm{p},z}$ in the fake calculation.
We therefore attribute the presence of a non-zero $\sigma_{\mathrm{p},z}$ 
to interactions with other atoms in the crystal, which
are neglected in the simple \mbox{B-O-B} picture. 

\begin{figure*}[p]
\begin{center}
\includegraphics[scale=0.30]{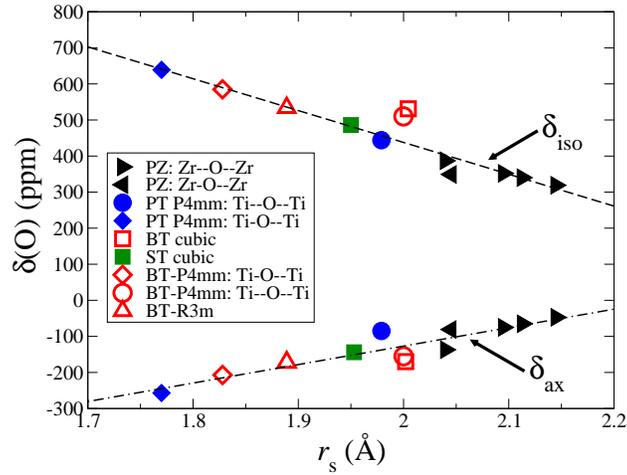}
\caption{Calculated oxygen isotropic $\delta_\mathrm{iso}$ and uniaxial
$\delta_\mathrm{ax}$ chemical shifts
in SrTiO$_3$ (ST), BaTiO$_3$ (BT), PbTiO$_3$ (PT), PbZrO$_3$ (PZ),
as a function of $r_\mathrm{s}$, the shortest B-O bond length of the targeted
O atom.
The
notation \mbox{B\,-\,-\,O\,-\,-\,B} indicates O atoms with two equidistant nn B atoms, 
and
\mbox{B\,-\,O\,-\,-\,B} indicates an O atom with one short and one long nn B bond ({\em e.g.}, 
along the polar c-axis in P4mm PbTiO$_3$). 
The straight lines are linear fits.
}
\label{fig:PT-PZplot}
\end{center}
\end{figure*}

Variations in \mbox{B-O-B} bond distances will affect O(2p)-B($n$d) hybridization,
and this should be reflected by corresponding
changes in the calculated oxygen chemical shieldings.
This is the case, as seen in
Figure~\ref{fig:PT-PZplot}, which shows 
that 
the isotropic 
$\delta_\mathrm{iso}$ and uniaxial
$\delta_\mathrm{ax}$ chemical shifts [Eq.(~\ref{eq:sigmaiso})]
exhibit a remarkably linear variation
as a function of $r_\mathrm{s}$, the shortest B-O bond length of the targeted
O atom. \cite{B-O_comment} Indeed, $\delta_\mathrm{iso}$ changes by more than
a factor of two over the plotted range. 
%

\begin{figure*}[p]
\begin{center}
\includegraphics[scale=0.30]{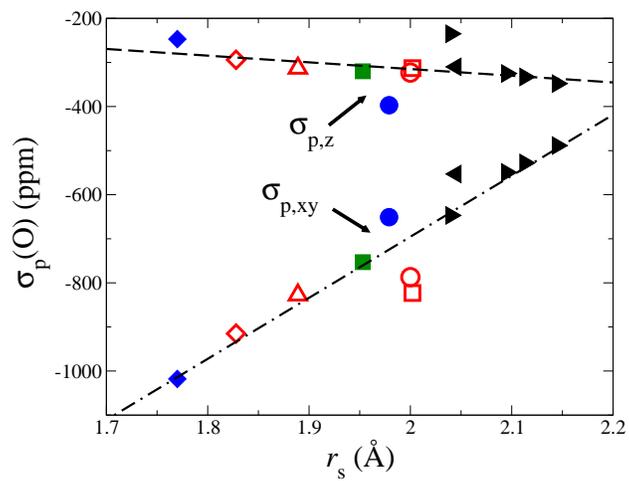}
\caption{
Calculated 
paramagnetic $\sigma_{\mathrm{p},z}$ and $\sigma_{\mathrm{p},xy}$ components (see text),
as a function of $r_\mathrm{s}$, the shortest B-O bond length of the targeted
O atom.
The straight lines are linear regressions of the points.
Symbols as in Fig.~\ref{fig:PT-PZplot}.
}
\label{fig:paramag}
\end{center}
\end{figure*}

The linear behavior is largely due to the deshielding $\sigma_x$ and $\sigma_y$
components in Table~\ref{tab:O-tensors}. 
To frame the discussion in terms of the paramagnetic components of the shielding tensors,
we first subtract the diamagnetic component of the isolated $\mathrm{O}^{2-}$ atom,
$\sigma(\mathrm{O}^{2-})=410$~ppm, defining 
$\sigma_{\mathrm{p},z}=\sigma_z-\sigma(\mathrm{O}^{2-})$ 
and 
$\sigma_{\mathrm{p},xy}=(\sigma_x+\sigma_y)/2-\sigma(\mathrm{O}^{2-})$.
These are plotted in Fig.~\ref{fig:paramag}. 
An alternative choice would be to plot the GIAO paramagnetic component. 
Even had we done GIAO calculations for all the different O-sites, this choice 
is not necessarily better, since both the diamagnetic and paramagnetic GIAO components
include contributions from neighboring atoms. \cite{Ditchfield1974} Moreover,
the separation into diamagnetic and paramagnetic components is, at best, only qualitatively
useful, since the separate components depend on the choice of gauge method.
The present choice is physically motivated, using a well defined diamagnetic reference system.
As indicated by the SrTiO$_3$ GIAO diamagnetic components, which 
differ by only $\simeq 40 - 50$~ppm from the O$^{2-}$ value (Table~\ref{tab:O-tensors}),
this subtraction largely removes the closed shell diamagnetic response.
In any case, our definition of $\sigma_{\mathrm{p}}$
simply shifts all $\sigma$ principal values by a constant.
As seen in Fig.~\ref{fig:paramag}, both $\sigma_{\mathrm{p},z}$ 
and $\sigma_{\mathrm{p},xy}$ 
vary linearly with $r_\mathrm{s}$.
The average slope of $\sigma_{\mathrm{p},xy}$ is $\simeq 5$ times larger
in magnitude than that of $\sigma_{\mathrm{p},z}$. This shows that the linear behavior
in Fig.~\ref{fig:PT-PZplot} is largely due to the variations of the deshielding $\sigma_x$ and $\sigma_y$
components, which arise from the p-d hybridization mechanism.

The p-d hybridization mechanism also plays a key
role in producing the anomalously large dynamical (Born) effective charge tensors $Z^*$,
which are universally seen in perovskite ferroelectrics for the O, transition metal B, 
and Pb atoms. 
\cite{BornZ-KingSmith,1993-Berryphase,1994-BerryRMP,Z-KNbO31996,BornZ-Ghosez-1998} 
In cubic BaTiO$_3$, SrTiO$_3$, PbTiO$_3$ and PbZrO$_3$, $Z^*_{zz}$(O)
(the component along the Ti-O bond) 
takes on the values
$-5.59$, $-5.66$, $-5.83$, and $-4.81$\cite{BornZ-KingSmith}, respectively, 
which is much larger than the nominal -2 value for O$^{2-}$. 
By contrast, the perpendicular components $Z^*_{xx}$(O)=$Z^*_{yy}$(O) are given by
$-2.11$, $-2.00$, $-2.56$, $-2.48$, respectively, which are much closer to the nominal value,
since these stretch the B-O bond only in second order.
Similarly, all $Z^*(\rm B^{4+})\simeq 6-7$ are anomalously large, 
since these involve the flow of dynamic charge as the B-O bond is stretched or compressed.
[In PbTiO$_3$ and PbZrO$_3$, $Z^*(\rm Pb)\simeq 4$ is anomalously large compared
to $Z^*(\rm Ba) \simeq Z^*(\rm Sr) \simeq 2.6$,
due to strong Pb-O covalency, and this is reflected in 
the somewhat more anomalous $Z^*_{xx}$(O) and $Z^*_{yy}$(O) in the Pb based crystals.]
Artificially decreasing the p-d hybridization, as in \Ref{Posternak1994}, resulted
in nominal values of all the $Z^*$. 

Similarly, the fake SrTiO$_3$ calculation eliminates B(d)-O(2p)
paramagnetic contributions to $\hat{\sigma}$ and largely removes the anisotropy
between $\sigma_{x}\simeq \sigma_{y}$ and $\sigma_{z}$ components.
B(d)-O(2p) covalency is essential for ferroelectricity in transition metal perovskites,
mitigating the repulsion between otherwise rigid ion-cores, which tends to suppress ferroelectric distortion;
under pressure, this mechanism eventually fails and ferroelectricity disappears. \cite{CohenKrak:1990}
Figure~\ref{fig:paramag} shows that the 
the anisotropy between $\sigma_{\mathrm{p},xy}$ and $\sigma_{\mathrm{p},z}$ increases linearly
with decreasing $r_s$, consistent with this picture, and
indicating a strengthening of the p-d hybridization mechanism
as $r_s$ is reduced.

\subsection{Exchange and correlation effects}

Despite the fact that the B3LYP band gap is nearly a factor of two larger than that of GGA,
the results in Tables~\ref{tab:O-tensors} and
\ref{tab:ST-BTshift} are fairly similar, with the exception of PbTiO$_3$,
where the B3LYP results are in better agreement with experiment (Table~\ref{tab:PT-PZshift};  
Sec.~\ref{sec:discuss-O-vs-exp}).
Although larger band gaps result in larger energy denominators
in the paramagnetic perturbative equations, \cite{Ramsey1950,Ramsey1951}
there is little difference in the shieldings between the two methods.
Similarly, a significant deshielded SrTiO$_3$ RHF paramagnetic component is evident
in Table~\ref{tab:O-tensors},
despite the much larger RHF HOMO-LUMO gap of 8.8 eV, compared to the B3LYP 3.6 eV gap.
%
The apparent dependence of the chemical shielding on the band gap can be somewhat misleading, however,
since the transverse paramagnetic component of the first order current density
could be made to vanish under suitable gauge transformations within the CSGT gauge method. \cite{ref:CSGT,zanasi1995} 
The entire shielding would then be given by the diamagnetic component, which depends only on the
occupied single particle states.
Similar observations were made regarding
calculations of the the spontaneous polarization $P$ and $Z^*$ in ferroelectric perovskites,
where RHF results for these quantities
were
found to be in good numerical agreement with both experiment and with
DFT calculations, \cite{HF-Polarization,HF-Born-Z} 
both methods yielding anomalously large $Z^*$.

\subsection{Comparison of oxygen chemical shifts with experiment}
\label{sec:discuss-O-vs-exp}

The calculated results in Table~\ref{tab:ST-BTshift} for SrTiO$_3$ and
BaTiO$_3$ are in very good agreement with single crystal
measurements. Overall, both B3LYP and GGA yield similar agreement with
experiment, while RHF is significantly worse.  
Comparisons for the
cubic and ferroelectric P4mm tetragonal BaTiO$_3$ phases should keep in mind
evidence for static and/or motional disorder of local structures suggested by 
experiment \cite{Comes1968,XAFS-disorder-1996,ref:BT-ST-O_Blinc} and theory,
\cite{Chain-KNbO3-PRL:1995,Krak:cond-mat97} 
since the present calculations were performed for 
ordered crystalline structures.

In PbTiO$_3$, the  O$_{\rm ax}$ atom has two nn Ti atoms along the polar direction, with one short
1.77~{\AA} and one long 2.39~{\AA} bond length, while the O$_{\rm eq}$ has
two equal bonds of 1.98~{\AA}. 
The experimental assignment of the two measured peaks is simplified by the fact that
the relative integrated intensities of the two O spectral peaks corresponds
to the 1:2 ratio of the O$_{\rm eq}$ to O$_{\rm ax}$ atoms in the simple 5-atom primitive
unit cell. \cite{ref:Baldwin05}
The large $\simeq 200$~ppm experimentally observed \cite{ref:Baldwin05}
splitting between the O$_{\rm ax}$ and O$_{\rm eq}$ reflects their very different Ti-O bond lengths.
The B3LYP calculated chemical shifts in Table~\ref{tab:PT-PZshift} reproduce
the $\simeq 200$~ppm splitting, while GGA underestimates it by $\simeq 50$~ppm.
Basis set and cluster size convergence errors (Section~\ref{sec:QMchemistry}) can be expected to
largely cancel in these calculated splittings, 
reducing the residual uncertainty to less than about $10$~ppm.
We are not aware of any measurements of the $^{17}$O uniaxial asymmetry in PbTiO$_3$.
 
In PbZrO$_3$ the peak assignments are more difficult, because four of the five
inequivalent oxygen sites have the same ratio of occurrence in the unit cell.
In \Ref{ref:Baldwin05} the peak assignment in the NMR spectrum was made based on the 
assumptions that (1) 
the proximity of the Zr cation plays the most important 
shielding role and (2) that the largest chemical shift corresponds to the largest 
bond length. The present calculations indicate that the first assumption is correct, but not 
the second. Instead, as seen in Fig.~\ref{fig:PT-PZplot}, 
the largest chemical shift corresponds to 
shortest bond distance between the targeted O atom and its nearest B atom.
Indeed, the figure shows that this holds true, not only for PbZrO$_3$,
but across several compounds, over a wide range of chemical shifts.
Our calculations indicate that chemical shift site assignments for 
the O(4) and O(5) atoms in \Ref{ref:Baldwin05} (their Table 1)  should be reversed. 
Finally, we note that differences between theory and experiment in the relative splittings are smaller
than those of the absolute splittings, {\em i.e.}, a rigid shift of the calculated B3LYP (GGA)
values in Table~\ref{tab:PT-PZshift} by $\simeq 10$~ppm ($\simeq -15$~ppm) 
removes most of the the discrepancies.

\section{Summary}
\label{conclusions}

We have shown that first-principles embedded cluster calculations,
using the DFT hybrid B3LYP method, can accurately calculate O 
chemical shifts for prototypical perovskite structure transition metal
oxides.  
Calculated isotropic and uniaxial chemical shifts
were found to be in good agreement with recent single crystal NMR $^{17}$O
measurements for SrTiO$_3$ and BaTiO$_3$.
For ferroelectric P4mm BaTiO$_3$ and PbTiO$_3$, the calculations
accurately reproduced measured power spectra NMR isotropic chemical shifts $\delta_\mathrm{iso}$
for the two inequivalent O sites.
The large $\simeq 200$~ppm
experimental splitting in P4mm PbTiO$_3$ is well reproduced by B3LYP, but DFT/GGA
underestimates it by $\simeq 50$~ppm.
In PbZrO$_3$, experimental and calculated $\delta_\mathrm{iso}$ are in very good agreement,
but experimental peak assignments are more difficult, since four of the five
inequivalent O sites appear in the same ratio. Our calculations, 
indicate a correction of the experimental assignments in \Ref{ref:Baldwin05}.

Our most notable findings are 1) a large anisotropy in the 
chemical shielding tensor, between deshielded  $\sigma_{x} \simeq \sigma_{y}$  and shielded $\sigma_{z}$
components, the latter principal axis being along the Ti-O bond, and 
2) a nearly linear variation, across all the systems studied,  of
$\delta_\mathrm{iso}$ and $\delta_\mathrm{ax}$ as a function of
B-O-B bond asymmetry.
We have shown that the anisotropy and linear variation arise from large paramagnetic 
contributions to $\sigma_{x}$ and $\sigma_{y}$ due to virtual transitions between 
O(2p) and unoccupied B($n$d) states.
A qualitative explanation of the anisotropy was given and then confirmed by
calculations for a fake material with no d-states.

We have shown that O NMR chemical shifts are a sensitive indicator
of the local structure in perovskites with transition metal B-site
atoms, due to 
covalent O(2p)-B($n$d) interactions.
This indicates that $^{17}$O NMR
spectroscopy, coupled with first principles calculations, can be 
an especially useful tool
to study the local structure in complex perovskite alloys.  

\section{Acknowledgments}

This work was supported by ONR grants N00014-08-1-1235 and N00014-09-1-0300.
DLP acknowledges support from a Virginia Space 
Grant Consortium Graduate Research Fellowship. Calculations were 
performed at the National Energy Research Scientific
Computing Center (NERSC) and the Center for Piezoelectric by Design. We acknowledge useful
discussions with Gina Hoatson and Robert L. Vold.

\bibliography{paper} 

\begin{thebibliography}{77}
\bibitem[{B-O()}]{B-O_comment}
\bibinfo{note}{The short B-O bond length $r_\mathrm{s}$ can be expressed as
  $r_\mathrm{s}=r_\mathrm{avg}-\Delta/2$, where $r_\mathrm{avg}$ is the average
  of the short and long B-O distances and $\Delta$ describes the asymmetry of
  the B-O-B chain. Since $r_\mathrm{avg} \simeq 2$~{\AA} across these systems,
  $r_\mathrm{s}$ is also a measure of the dimerization or bond length asymmetry
  of the B-O-B chain.}
\end{thebibliography}

\begin{table}[p]
\begin{center}
\caption{ Calculated oxygen chemical shielding principal values (in
    ppm) for SrTiO$_3$, BaTiO$_3$, PbTiO$_3$ and PbZrO$_3$.  In all
    cases, ``$z$'' denotes the principal axis that is most closely
    along the B-O bond direction. Unless otherwise indicated, the CSGT
    gauge method was used, with B3LYP for exchange and correlation.
    For some cases, GGA results are given in parenthesis.  RHF and
    GIAO gauge method B3LYP results are given for SrTiO$_3$.  GIAO
    results for the isolated diamagnetic O$^{2-}$ atom are also shown.
    See the text for a discussion of the fake SrTiO$_3$
    calculation. For PbZrO$_3$, we adopt the site labeling convention
    of \Ref{ref:Johannes}; the O(i) site notation of
    \Ref{ref:Baldwin05} is also given in brackets.  }
\begin{tabular*}{0.48\textwidth}{@{\extracolsep{\fill}}llccc} \hline \hline 
~~~ &\\
 & & $\sigma_x$ & $\sigma_y$ & $\sigma_z$ \\[1pt]
\hline \\
\multicolumn{5}{c}{SrTiO$_3$(cubic)}    \\[1pt] 
         &  &-343 (-353)& -343 (-353)& ~90 (~46)\\ [1pt]
GIAO diamag.       &  &  347 & 347 & 371 \\ [1pt]
GIAO paramag.      &  & -694 & -694 & -255 \\ [1pt]
GIAO total      &  &  -347 & -347 & 116 \\ [5pt]
 RHF   &  &-134 & -134 & 201 \\ [1pt]
 O$^{2-}$      &   &  410  & 410  & 410 \\ [5pt]
fake calculation&\\[1pt]
GIAO diamag.     &  &  397 & 397 & 448 \\ [1pt]
GIAO paramag.  &  & -228 & -228 & -262 \\ [1pt]
GIAO total   &  &  169 & 169 & 186 \\ [1pt]
\hline \\
\multicolumn{5}{c}{BaTiO$_3$}  \\[1pt] 
cubic &   &-413 (-414)& -413 (-414)& ~97 (49)\\[1pt]
P4mm & O$_{\rm ax}$ &-505 (-480)& -505 (-480)& 116 (70)\\
     & O$_{\rm eq}$ &-401 (-407)& -353 (-355)&  ~87 (40)\\ [1pt]
R3m &  & -423 (-416)& -412 (-406)& ~97 (51)\\[1pt]
\hline \\
\multicolumn{5}{c}{PbTiO$_3$(P4mm)}   \\[1pt] 
& O$_{\rm ax}$   &  -608 (-562)& -608 (-562) & 163 (123)\\
     & O$_{\rm eq}$   &  -263 (-286)& -219 (-228)&  ~13 (-32)\\ [1pt]
\hline \\
\multicolumn{5}{c}{PbZrO$_3$(Pbam)}   \\[1pt] 
\multicolumn{2}{l}{O1-4g [O(1)]}                   & -172 (-197)& -107 (-130)& ~85 (~48)\\
\multicolumn{2}{l}{O1'-4g [O(2)]}                  & -132 (-158)& -104 (-126)& ~77 (~40)\\
\multicolumn{2}{l}{O2-8i [O(3)]}                   & -147 (-171)& -139 (-162)& 100 (~64)\\
\multicolumn{2}{l}{O3-4f [O(4)]\footnotemark[1]}           & ~-93 (-118)& ~-64 (~-90)& ~62 (~26)\\
\multicolumn{2}{l}{O4-4e [O(5)]}                   & -251 (-269)& -223 (-239)& 175 (136)\\[2pt]
\hline \hline
\end{tabular*}
\footnotetext[1]{{Note: our calculations suggest that the assignment of the O3-4f[O(4)] and O4-4e[O(5)] peaks should be reversed in Ref.\ \onlinecite{ref:Baldwin05}, as done here.}}
\vspace{-10pt}
\label{tab:O-tensors}
\end{center}
\end{table}

\begin{table*}[p]
\caption{
Comparison between calculated and measured single crystal and powder spectra 
$^{17}$O chemical shifts (in ppm) for
BaTiO$_3$ and SrTiO$_3$. 
Calculated isotropic shifts $\delta_{\rm iso}$ and uniaxial components $\delta_{\rm ax}$ are derived
from Table~\ref{tab:O-tensors}, referenced to
liquid water,  $\sigma_\mathrm{iso}^{\mathrm{water}}=287.5$~ppm (\Ref{ref:wasylishen}).
Calculated values are from B3LYP (GGA values
are in parenthesis), and RHF results are also shown for SrTiO$_3$. 
} 
\begin{tabular*}{0.98\textwidth}{@{\extracolsep{\fill}}llcccc} \hline \hline 
~~~ &\\
& & \multicolumn{2}{c}{$\delta_{\rm iso}$} & \multicolumn{2}{c}{$\delta_{\rm ax}$} \\ 
& &  Theory &  Expt.&  Theory &  Expt. \\[3pt]
\hline
\\
SrTiO$_3$&\\[5pt]
cubic &      & 486 (507) & 467 $\pm$ 5\footnotemark[1], 465\footnotemark[2]   & -144 (-132) & -135.3 $\pm$ 5\footnotemark[1]     \\ 
cubic RHF                &   & 310 &  & -112 &                  \\
\\
\hline 
~~~ &\\
BaTiO$_3$&\\[5pt]
cubic  &     & 530 (547)& 546 $\pm$ 5\footnotemark[1]  & -170 (-154) & -150 $\pm$ 1\footnotemark[1]  \\   
$\rm P4mm$& O$_{\rm ax}$  & 585 (584)& 570 $\pm$ 5\footnotemark[1], 553\footnotemark[3], 564\footnotemark[4]  & -207 (-183) & -171 $\pm$ 1\footnotemark[1]  \\ 
          & O$_{\rm eq}$  & 510 (528)& 520 $\pm$ 5\footnotemark[1], 530\footnotemark[3], 523\footnotemark[4]  & -155 (-140)& -142 $\pm$ 1\footnotemark[1]  \\  [5pt]

\hline \hline
\end{tabular*}
\footnotetext[1]{Single crystal experimental values are from Ref.\ \onlinecite{ref:BT-ST-O_Blinc}}
\footnotetext[2]{Powder results Ref.\ \onlinecite{ref:BT-O_Bastow}}
\footnotetext[3]{Powder results Ref.\ \onlinecite{ref:BT_Stebbins}}
\footnotetext[4]{Powder results Ref.\ \onlinecite{ref:BT-O_Anuradha}}
\label{tab:ST-BTshift}
\end{table*}

\begin{table}[p]
\vspace{-0.1in}
\caption{Comparison between calculated and experimental isotropic chemical shifts 
$\delta_{\rm iso}$ (in ppm) for PbTiO$_3$ and PbZrO$_3$.  
Calculated isotropic shifts are derived
from Table~\ref{tab:O-tensors}, referenced to
liquid water,  $\sigma_\mathrm{iso}^{\mathrm{water}}=287.5$~ppm (\Ref{ref:wasylishen}).
Calculated values are from B3LYP (GGA values
are in parenthesis). For PbZrO$_3$, we adopt the site labeling convention
of \Ref{ref:Johannes}; the O(i) site notation of \Ref{ref:Baldwin05}
is also given.
}
\begin{tabular*}{0.48\textwidth}{@{\extracolsep{\fill}}lcc} \hline \hline 
~~~ &\\
& Theory & Expt.\footnotemark[1] \\[3pt]
\hline
\\
PbTiO$_3$&\\[5pt]
O$_{\rm ax}$       & 639 (621)& 647 $\pm$ 2   \\
O$_{\rm eq}$       & 444 (469)& 443 $\pm$ 2   \\ 
\\
\hline
~~~ &\\
PbZrO$_3$&\\[5pt]
O1-4g [O(1)]       & 352 (380)& 365 $\pm$ 2    \\
O1'-4g [O(2)]       & 340 (369)& 351 $\pm$ 2    \\
O2-8i [O(3)]       & 349 (377)& 356 $\pm$ 2    \\
O3-4f [O(4)] \footnotemark[2]      & 319 (348)& 329 $\pm$ 2    \\ 
O4-4e [O(5)]       & 387 (412)& 408 $\pm$ 2    \\[5pt]
\hline \hline
\end{tabular*}
\footnotetext[1]{Experimental values are from Ref.\ \onlinecite{ref:Baldwin05}}
\footnotetext[2]{Note: our calculations suggest that the assignment of the O3-4f[O(4)] and O4-4e[O(5)] peaks should be reversed in 
Ref.\ \onlinecite{ref:Baldwin05}, as done here.}
\label{tab:PT-PZshift}
\end{table}

\end{document}